\pdfoutput=1
\documentclass[11pt]{article}

\usepackage[final]{acl}

\usepackage{times}
\usepackage{latexsym}

\usepackage[T1]{fontenc}
\usepackage[utf8]{inputenc}

\usepackage{microtype}

\usepackage{inconsolata}

\usepackage{graphicx}

\usepackage{booktabs}
\usepackage{amsfonts}

\title{Streaming Speaker Change Detection and Gender Classification for Transducer-Based Multi-Talker Speech Translation}

\author{\textbf{Peidong Wang}\textsuperscript{}, \textbf{Naoyuki Kanda}\textsuperscript{}, \textbf{Jian Xue}\textsuperscript{}, \textbf{Jinyu Li}\textsuperscript{}, \textbf{Xiaofei Wang}\textsuperscript{}, \\ \textbf{Aswin S. Subramanian}\textsuperscript{}, \textbf{Junkun Chen}\textsuperscript{}, \textbf{Sunit Sivasankaran}\textsuperscript{}, \textbf{Xiong Xiao}\textsuperscript{}, \textbf{Yong Zhao}\textsuperscript{} \\
        \\
        \textsuperscript{}Microsoft \\ 
        One Microsoft Way, Redmond WA, USA}

\begin{document}
\maketitle
\begin{abstract}
Streaming multi-talker speech translation is a task that involves not only generating accurate and fluent translations with low latency but also recognizing when a speaker change occurs and what the speaker's gender is. Speaker change information can be used to create audio prompts for a zero-shot text-to-speech system, and gender can help to select speaker profiles in a conventional text-to-speech model. We propose to tackle streaming speaker change detection and gender classification by incorporating speaker embeddings into a transducer-based streaming end-to-end speech translation model. Our experiments demonstrate that the proposed methods can achieve high accuracy for both speaker change detection and gender classification.
\end{abstract}

\section{Introduction}
\label{sec:intro}
Speech translation (ST) aims to convert spoken words in one language to text in another language with high accuracy and fluency. A common approach is to use two components: an automatic speech recognition (ASR) system that converts the speech input into text, and a machine translation (MT) system that translates the text into the target language~\cite{Ney1999ST, Matusov2005ST, Post2013ST}. This method is commonly referred to as cascaded ST.

To avoid error accumulation and fully exploit the audio information, end-to-end (E2E) ST was studied~\cite{vila2018end,sperber2020speech,li2022recent}, which uses a single model to translate directly from speech to text. 
The most well-studied architecture for E2E ST is attention-based encoder-decoder (AED) models~\cite{wang2019token,wang2019large,wang21t_interspeech}, which use an attention layer to combine audio representations and text.
The authors of~\cite{Berard2016ST} suggest using AED models~\cite{chan2015listen,wang21t_interspeech} for a small French-English dataset. A similar model architecture is used in~\cite{weiss2017sequence} for the Fisher Callhome Spanish-English task and achieves better results than the cascaded approach on the Fisher test set. AED-based models are also applied in~\cite{Berard2018ST} for a large-scale E2E ST task. However, AED models usually work in an offline mode that requires the whole utterance to be available before decoding can begin.
% It is difficult to design the streaming strategy for such models.
% It is thus difficult to design the streaming strategy for such models. 
Monotonic chunkwise attention (MoChA)~\cite{chiu2018monotonic} is an attention mechanism that enables efficient and online decoding. Several variants of MoChA have been developed, such as monotonic infinite lookback attention (MILk)~\cite{arivazhagan2019monotonic}, monotonic multi-head attention~\cite{ma2019monotonic, ma2021streaming}, multitask learning~\cite{miao2019online}, and minimum latency training strategies~\cite{inaguma2020minimum}. These methods aim to enhance the performance and robustness of MoChA in various domains and scenarios.
% E2E ST and ASR are similar in that they are both sequence-to-sequence mappings. Many model architectures can thus be shared, especially between ST using monotonic alignments~\cite{raffel2017online} and ASR. To enable more effective communication between users, streaming (i.e., simultaneous) models are topics of investigation in both areas. Monotonic chunkwise attention (MoChA)~\cite{chiu2018monotonic} was used in both MT and ASR. The MT version was extended to monotonic infinite lookback attention (MILk)~\cite{arivazhagan2019monotonic} and monotonic multi-head attention~\cite{ma2019monotonic, ma2021streaming}, and the ASR version was improved by multitask learning~\cite{miao2019online} and minimum latency training strategies~\cite{inaguma2020minimum}. Another streaming model architecture is the neural transducer~\cite{prabhavalkar2017asr,sainath2020asr,li2020asr,saon2021asr}, which outperforms MoChA and has emerged to be the state-of-the-art (SOTA) streaming E2E model in ASR~\cite{li2021recent}, but has been less investigated in ST. Recently, Liu \emph{et al.} proposed cross attention augmented transducer (CAAT) for ST~\cite{liu2021caat}. It uses Transformers in the joint network to combine encoder and prediction network outputs. Due to the use of Transformers and multi-step decision for memory footprint reduction, the latency of CAAT is large. In addition, to train a CAAT, complicated regularization terms and extensive hyper-parameter tuning are required.

Recently, transducer-based models have emerged as promising candidates for end-to-end speech translation (E2E ST)~\cite{Graves-RNNSeqTransduction,prabhavalkar2017asr,sainath2020asr,li2020asr,saon2021asr}, especially in streaming scenarios.
Streaming ST is important for real-time communications, where we do not want to wait until the end of a sentence to perform translation~\cite{papi2024real}.
The transducer loss accounts for all the potential paths from the source audio to the target texts, which may improve the model's convergence.
The authors of~\cite{xue2022large} proposed to use neural transducers for ST, which was later extended to a many-to-many ST and ASR model~\cite{wang2022lamassu} and applied to large-scale datasets with true zero-shot capability~\cite{xue2023weakly}. 
Many methods were proposed to combine AED and neural transducer models. Liu \emph{et al.} proposed cross-attention augmented transducer networks for streaming ST~\cite{liu2021caat}. Tang \emph{et al.} proposed hybrid transducer and attention-based encoder-decoder modeling~\cite{tang2023hybrid}.

One of the main difficulties of applying ST to real-world situations is dealing with multiple talkers~\cite{wang2019speech,wang2020speaker,wang2018utterance}. In video dubbing tasks, the ST models should not only transcribe the speech, but also identify the speaker change points and their genders. In this way, a text-to-speech (TTS) module can generate realistic and consistent audio output. Speaker change detection is particularly important for the state-of-the-art zero-shot TTS models, which rely heavily on the quality of the audio prompt. For conventional TTS models using speaker profiles, gender classification is required to ensure that the selected speaker profile has the correct gender. The speaker change segmentation and the ST text output should be synchronized. This means that existing speaker diarization methods such as 
% Whisper~\cite{radford2023robust} and 
speaker clustering-based method~\cite{park2021review} or neural speaker diarization (e.g., EEND~\cite{fujita2019end,fujita2019end2}) may not be easily applied to this task.

Only a few studies have been conducted on this new challenge. Zuluaga-Gomez \emph{et al.}~\cite{zuluaga2023end} investigated speaker diarization for offline ST based on an AED model. Moreover, Yang \emph{et al.}~\cite{yang2023diarist} proposed t-SOT~\cite{kanda22arxiv} and t-vector~\cite{kanda2022streaming} based streaming multi-talker ST methods, where they evaluated the model based on the speaker diarization task. 

% This paper aims to tackle this new challenge. 
In this paper, we focus on streaming speaker change detection and gender classification, which are essential for streaming speech-to-speech translation models that have an ST frontend and a TTS backend. We note that our methods are not limited to English-to-many (EN-to-many) translation, but can also be applied to many-to-EN translation. For streaming speaker embedding generation, we used the t-vector method~\cite{kanda2022streaming}, which was originally proposed to produce a token-wise speaker embedding for multi-talker speech recognition. 
A speaker change detection and gender classification are performed on the estimated t-vector, which are evaluated with multiple language pairs.
Our work is closely related to that of Yang \emph{et al.}~\cite{yang2023diarist}, with a greater focus on speaker change detection and gender detection, areas that have not been previously investigated.
The remainder of this paper is organized as follows. Section \ref{sec:method} presents the proposed method. Section \ref{sec:exp} describes the experimental setup. Section \ref{sec:eval} discusses the results and their implications. Section \ref{sec:conc} summarizes the main contributions.

\section{Method}
\label{sec:method}
First, we introduce multilingual speech translation based on transducers, a method that achieves high-quality streaming speech translation. We then describe the t-vector method. Finally, we show how we combined these two methods for streaming speaker change detection and gender classification.

\subsection{Transducer-based streaming multilingual ST}
\label{ssec:trans_streaming_multilingual}

% We adopt the LAMASSU-UNI approach to support English-to-many ST. It uses language ID as the initial token for the prediction network.
\subsubsection{Transformer transducer}
\label{sssec:trans}
\begin{figure}[ht]
    \centering
    \includegraphics[width=0.5\textwidth]{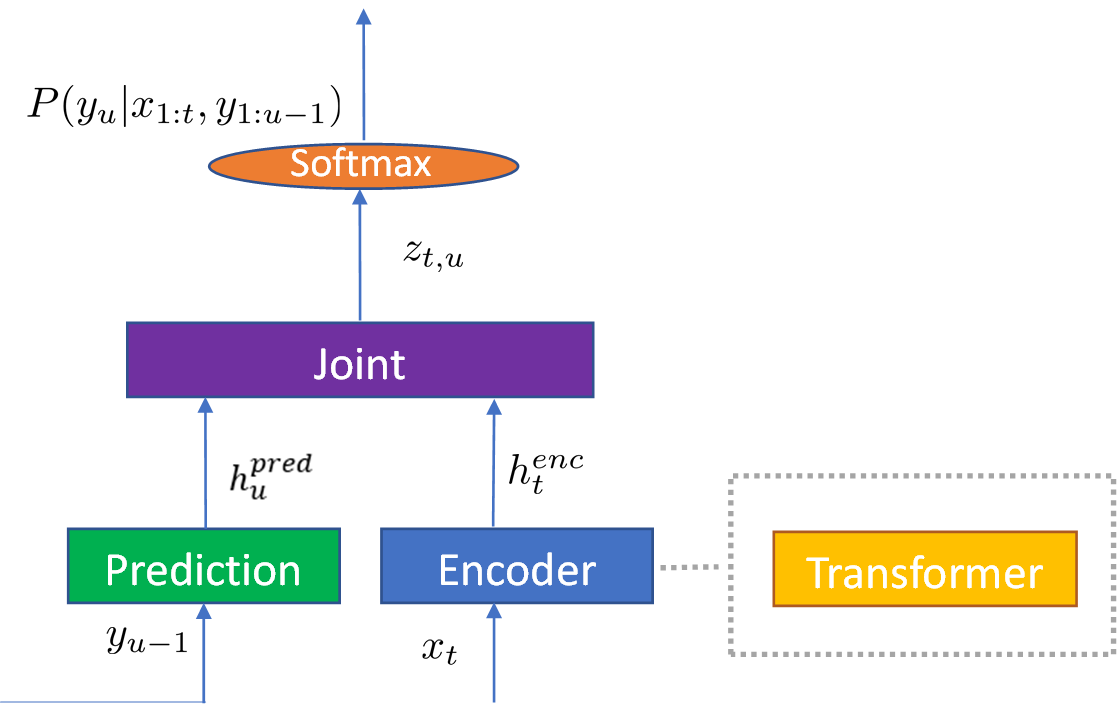}
    \caption{Illustration of Transformer transducer for ST.}
    \label{fig:T-T_fig}
\end{figure}

A transducer model consists of three components, as shown in Figure \ref{fig:T-T_fig}: an encoder, a prediction network, and a joint network. The encoder takes $d_x$-dimensional audio features $\textbf{x}_t \in \mathbb{R}^{d_x}$ as input and generates $d_e$-dimensional hidden states $\textbf{h}_t^{\mathrm{enc}} \in \mathbb{R}^{d_e}$. The prediction network uses the embedding of the previous non-blank output token $\textbf{y}_{u-1} \in \mathbb{R}^{1}$ to generate the hidden state $\textbf{h}_u^{\mathrm{pred}} \in \mathbb{R}^{d_p}$ for step $u$. The joint network combines $\textbf{h}_t^{\mathrm{enc}}$ and $\textbf{h}_u^{\mathrm{pred}}$ into a $T \times U$ matrix represented by $\textbf{z}_{t,u} \in \mathbb{R}^{d_z}$, and then applies a softmax function to obtain probabilities for paths that align audio frames with token sequences. The model uses a blank token at the output to handle the alignment between audio and text. During training, the model considers all possible paths and maximizes the probabilities of the correct paths.
In this study, we use Transformer transducer (T-T). It uses Transformer blocks in the encoder, which have a multi-head self-attention layer and a feedforward layer.

\subsubsection{Streaming capability}
\label{sssec:streaming}
\begin{figure}[ht]
    \centering
    \includegraphics[width=0.45\textwidth]{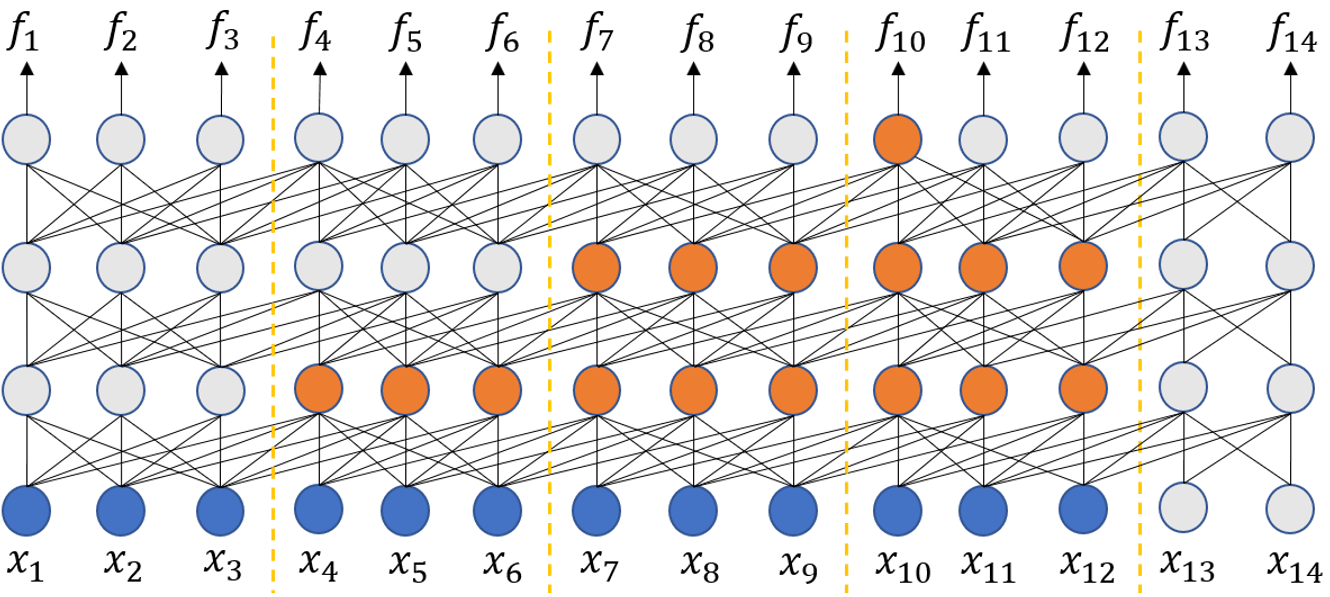}
    \caption{Illustration of the reception field of a streaming T-T at position $f_{10}$ with chunk size 3 and the number of left chunks 1.}
    \label{fig:reception_fig}
\end{figure}

The encoder receives the audio input in chunks to enable streaming. To stream with low latency and computation, an attention mask, as in~\cite{xiechen2021tt}, is used. At layer $l$, the input $\textbf{x}^l_{1:T}$ is split into chunks $\textbf{c}^l_{1:S}$ over time with chunk size $U$. At time step $t$, $\textbf{x}^l_t$ only attends to frames in its chunk $\textbf{c}^l_{t / U + 1}$ and $B$ left chunks $\textbf{c}^l_{\max(1, t / U + 1 - B):t / U}$. The reception field at each layer grows linearly with the number of layers, allowing the model to use a longer history for better performance, as shown in Figure \ref{fig:reception_fig}. The frames cannot see frames outside their chunk, keeping a fixed number of look-ahead frames.

\subsubsection{Multilingual capability}
\label{sssec:multilingual}
\begin{figure}[ht]
    \centering
    \includegraphics[width=0.36\textwidth]{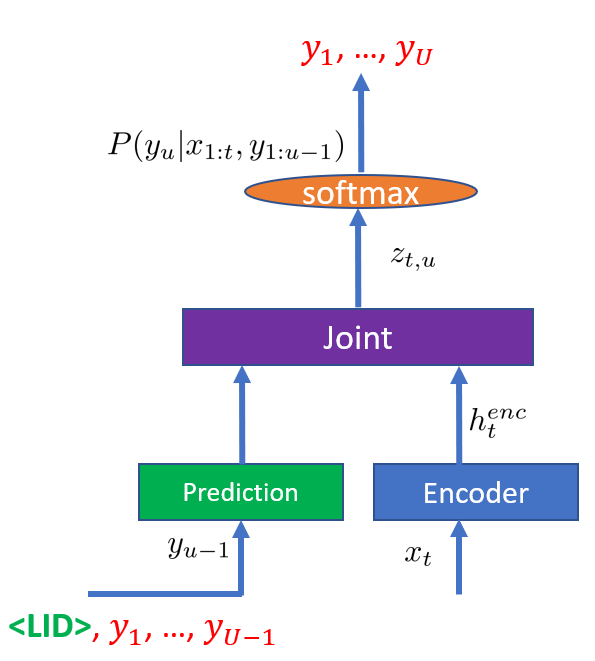}
    \caption{Illustration of LAMASSU-UNI.}
    \label{fig:multilingual_fig}
\end{figure}

We adopt the LAMASSU-UNI approach~\cite{wang2022lamassu}, which is illustrated in Fig \ref{fig:multilingual_fig}, to perform speech translation from English to various other languages, by providing the prediction network with a starting token that specifies the language. The language identifications (LIDs) are appended to the vocabulary list and are treated as special tokens.

% Fig. \ref{fig:transducer_single_decoder} shows LAMASSU using a unified prediction and joint network for multiple target languages (LAMASSU-UNI). During training, it uses target LID to replace the start of sentence token ($<SOS>$) in the input to the prediction network. At test time, target LID is fed to the prediction network as the initial token.

\subsection{t-vector for ST}
\label{ssec:tvec}
A t-vector is a type of speaker embedding vector that captures the speaker characteristics at the token level~\cite{kanda2022streaming}. It is based on the d-vector, which is obtained from audio segments. A t-vector module is typically added to a transducer model that has been well-trained. It was first developed for multi-talker ASR and has been recently used for ST tasks~\cite{yang2023diarist}.

\begin{figure}
    \centering
    \includegraphics[width=0.5\textwidth]{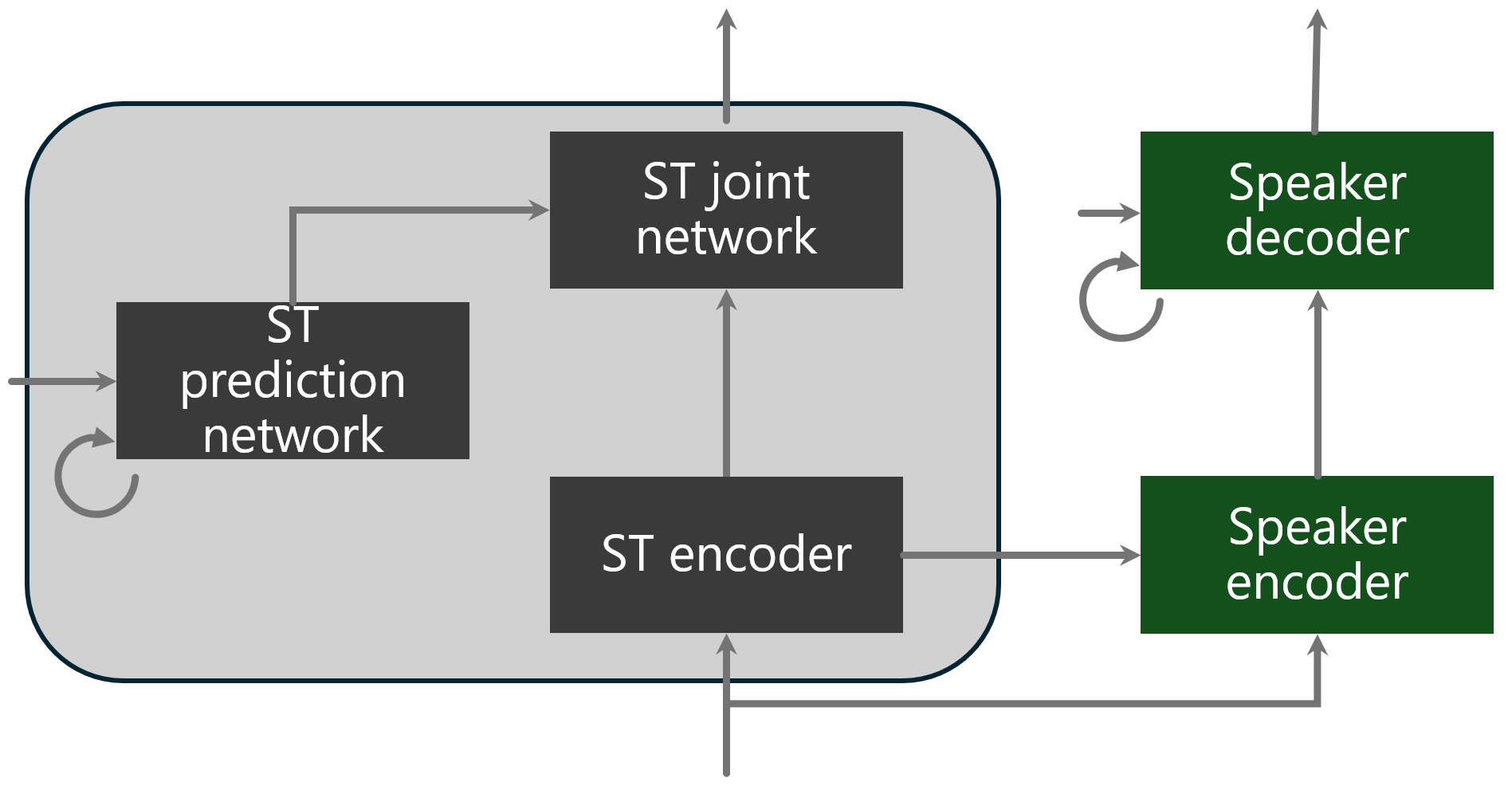}
    \caption{Illustration of t-vector model for ST.}
    \label{fig:t_vec}
\end{figure}

\begin{figure}[h]
    \centering
    \includegraphics[width=0.4\textwidth]{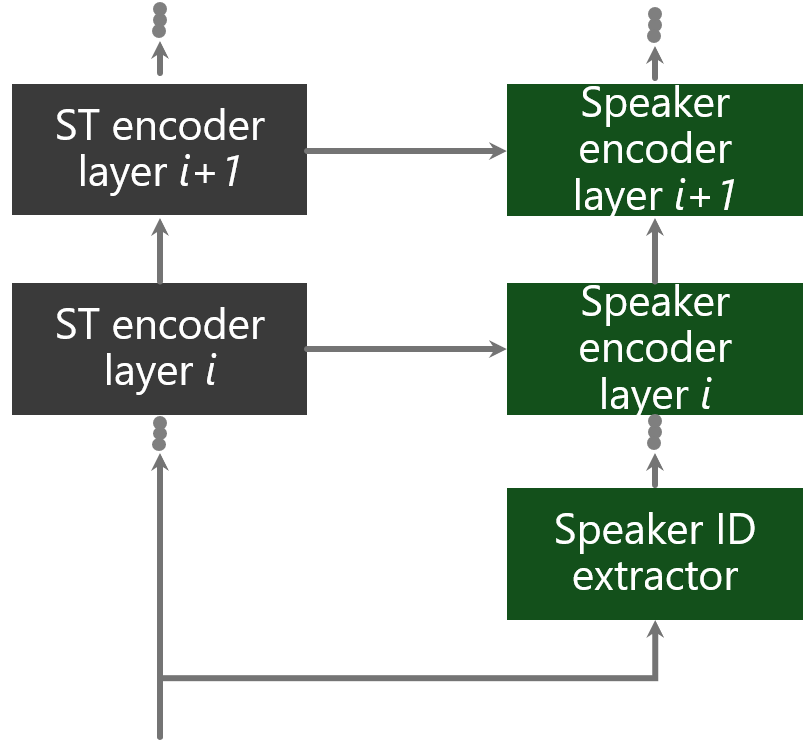}
    \caption{Illustration of the speaker encoder layers. The speaker ID extractor is typically a d-vector extractor.}
    \label{fig:speaker_encoder}
\end{figure}

The t-vector model, which is built on top of an ST model, is illustrated in Figure \ref{fig:t_vec}. The ST model is fixed during training, as depicted by the gray box. The t-vector module consists of a speaker encoder and a speaker decoder. The speaker encoder has multi-head attention layers that use external attention to extract the speaker information from the lower layer and the corresponding ST encoder layer. Specifically, the attention layers generate their own key and query from the lower layer, and use the output of the corresponding ST encoder layer as value. Figure \ref{fig:speaker_encoder} shows the details of the speaker encoder layers. The output of the speaker encoder, together with the embedding of the output non-blank token, is fed to the speaker decoder. The speaker decoder has two long short-term memory (LSTM) layers, whose output is passed through a linear layer to produce t-vectors.

For each output non-blank token of the ST model, t-vectors are generated in a streaming manner during inference. Unlike Yang \emph{et al.}~\cite{yang2023diarist}, we do not need speaker clustering for speaker diarization for this task, as our focus is on speaker change detection and gender classification. Moreover, since we did not change the ST part during t-vector training, the translation performance remains unaffected.

\begin{figure*}[ht]
    \centering
    \includegraphics[width=0.6\textwidth]{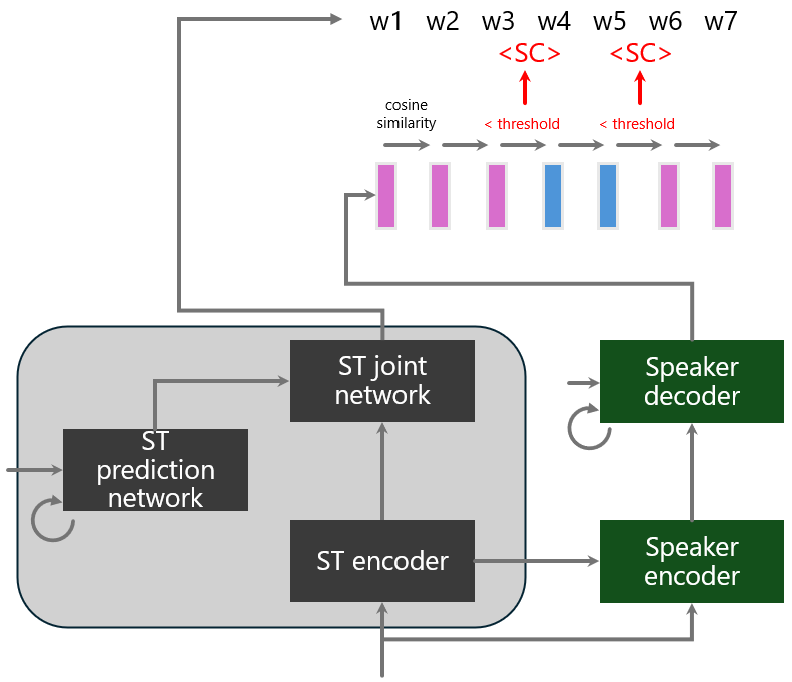}
    \caption{Illustration of speaker change detection using t-vectors. $w1$ to $w7$ denote non-blank output tokens. If the cosine similarity between adjacent tokens is below the threshold, we insert an $\langle\mathrm{SC}\rangle$ token to the output.}
    \label{fig:scd}
\end{figure*}

\subsection{t-vector for speaker change detection and gender classification}
\label{ssec:scd_gc}
In this section, we explain how we use transducer-based streaming multilingual ST and t-vector to perform streaming speaker change detection and gender classification.
Our model is data-efficient because it treats the speaker change detection problem as a speaker identification (SID) generation task. This means it does not rely on a large amount of real-world training data, which is difficult to acquire and may not cover all possible scenarios.

% We use these tasks to select the best audio segments for zero-shot TTS.
% Zero-shot TTS models require high-quality audio prompts to synthesize speech in various voices. To filter out the noisy or unsuitable prompts, our model performs two tasks: speaker change detection and gender classification.

\subsubsection{Speaker change detection}
\label{sssec:scd}
We use the cosine similarity between two adjacent t-vectors to detect a speaker change. If the similarity value is lower than a threshold, we assume that a different speaker is speaking. 
Our method does not require model retraining for the ST model part, and therefore can preserve the ST model performance. An alternative way to perform speaker change detection is to train a model with real conversations that have speaker changes. However, collecting such training data is challenging. If we use simulated data, the model tends to learn the channel variations between different audio segments rather than the speaker differences. Therefore, we choose the t-vector method for our study.

\subsubsection{Gender classification}
\label{sssec:gc}
% Correct gender classification ensures that the output audio from the TTS backend is distinguishable 
In conventional TTS systems, a specific speaker profile is chosen for each output audio segment. While these systems cannot retain all the audio nuances like zero-shot TTS, they are essential for on-device deployment to prevent malicious use of zero-shot TTS. To ensure natural-sounding conversations, it's crucial to accurately classify the gender of speakers for each audio segment; otherwise, listeners might become confused in understanding the translated conversations. Since we already produce t-vectors for detecting speaker changes, we aim to utilize this information for gender classification too. We begin by gathering a collection of speaker profiles categorized as male or female. We then calculate the cosine similarity between the t-vector of each token and the male and female speaker profiles. Lastly, we determine the gender of the token by selecting the gender with the highest cosine similarity score.
% We note that there are other methods for gender classification based on audio features, such as fundamental frequency detection, but they require aligning the translated text with the corresponding audio after processing the time-domain information. This is challenging for streaming ST tasks, where the output text is not monotonic. Therefore, our t-vector-based method is preferable because it operates on the token level.

\begin{figure*}[ht]
    \centering
    \includegraphics[width=0.82\textwidth]{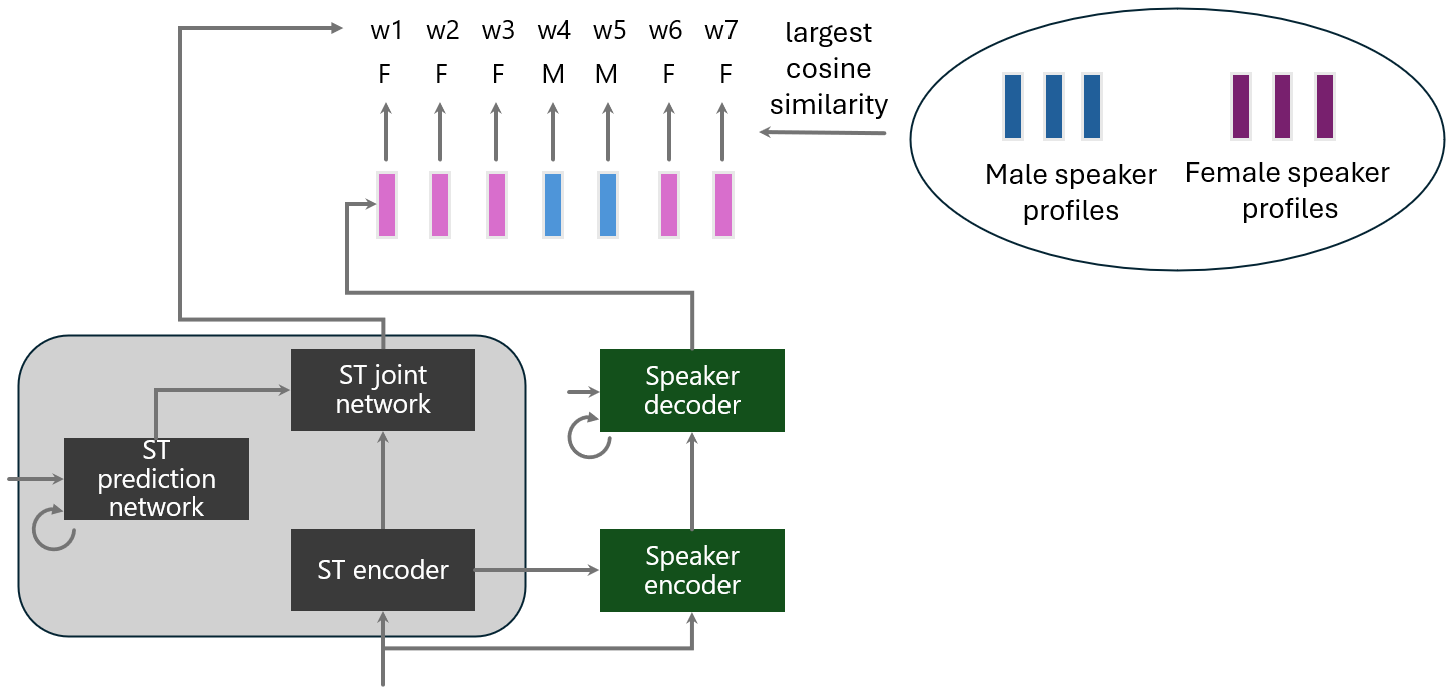}
    \caption{Illustration of gender classification using t-vectors. See the caption of Figure \ref{fig:scd} for the definitions of $w1$ to $w7$. The male and female speaker profiles contain d-vectors. $F$ and $M$ denote female and male, respectively. 
    % For each t-vector, we compare the cosine similarities with male and female speaker profiles, and assign $F$ or $M$ estimations corresponding to the largest cosine similarities between the two categories.
    }
    \label{fig:gc}
\end{figure*}

\section{Experimental setup}
\label{sec:exp}

\subsection{Data}
\label{ssec:exp_data}
\subsubsection{Training data}
\label{sssec:exp_data_training}
We have about 75K hours of English audio in the training data for the ST model. The audio is collected from various sources and is anonymized. For each of the five output languages Germany (DE), Spanish (ES), Hindi (HI), Italian (IT), and Russian (RU), we use text machine translation models to generate texts with pseudo-labels. 

We train the t-vector model using Voxceleb~\cite{nagrani2017voxceleb} as the training data and use our well-trained ST model to generate the label texts in each target language. We then apply a Viterbi algorithm to align the tokens and the audio frames. The t-vector training stage uses the alignment produced by this step to associate the t-vector with the non-blank tokens. This allows a cross-entropy training for the t-vector model that is memory-efficient and also produces the t-vector together with the output token during inference.

\subsubsection{Test data}
\label{sssec:exp_data_test}
Our test data consisted of 5 real recorded long audios. The test samples had an average duration of about 30 minutes.
We collected real conversational audios with various numbers of speakers, up to eight, for speaker change detection. Human annotators segmented the audio and marked the speaker change points at the segment boundaries. We concatenated the segments into audio samples, each containing one speaker change. The total number of samples was 688, with an average duration of about 5 seconds. Some samples were 30 seconds or longer.
For gender classification, we divided all the speaker profiles in the training set of Voxceleb-1~\cite{nagrani2017voxceleb} and Voxceleb-2~\cite{nagrani2020voxceleb} into two gender categories: male and female. We used the test set of Voxceleb-1 as the test audio. The total number of utterances we tested was 4874.

\subsection{Model}
\label{ssec:exp_model}
We applied a streaming multilingual T-T model that performs EN-to-many speech translation with a chunk size of 1s and 18 history chunks. The model has 12 Transformer encoder layers and about 100 million (M) parameters. The five output languages use a shared vocabulary of about 18587 tokens. We also added a language ID for each language to the vocabulary list. The total number of tokens, including $\langle\mathrm{EOS}\rangle$ and $\langle\mathrm{blank}\rangle$, is 18594.
We trained the model for 96M steps, with a peak learning rate of 3e-4 and 0.8M warm-up steps. We used fp16 in Deepspeed and 32 32G GPUs for training. For inference, we used PyTorch decoding with a beam size of 8.

The t-vector model consists of a speaker encoder and a speaker decoder. The speaker encoder uses a Res2Net module for SID extraction~\cite{yang2023diarist}. It has the same number of encoder layers as the ST encoder, which is 12 in our study. The attention dimension and the attention head are 128 and 8, respectively. The speaker decoder is a two-layer LSTM model with a hidden size of 512 and an input size of 128. The t-vector dimension is 128. We fix the ST model parameters during training. We also initialize the SID module inside the encoder with a pre-trained SID model, which is trained on the Voxceleb-1~\cite{nagrani2017voxceleb} and Voxceleb-2~\cite{nagrani2020voxceleb} training sets. The SID module remains fixed during training.

\section{Evaluation results}
\label{sec:eval}

\begin{table*}[]
    \centering
    \begin{tabular}{c c | c c c c c | c }
        \toprule
         threshold & metrics & EN-DE & EN-ES & EN-HI & EN-IT & EN-RU & Avg. \\
         \midrule
              & recall & 0.89 & 0.91 & 0.90 & 0.88 & 0.89 & 0.89 \\
         0.99 & precision & 0.55 & 0.55 & 0.55 & 0.55 & 0.55 & 0.55 \\
              & ${F_1}$ & 0.68 & 0.69 & 0.68 & 0.67 & 0.68 & 0.68 \\
         \midrule
              & recall & 0.63 & 0.67 & 0.63 & 0.67 & 0.65 & 0.65 \\ 
         0.94 & precision & 0.66 & 0.69 & 0.67 & 0.68 & 0.69 & 0.68 \\
              & ${F_1}$ & 0.65 & 0.68 & 0.65 & 0.67 & 0.67 & 0.66 \\
         \midrule
              & recall & 0.49 & 0.53 & 0.51 & 0.52 & 0.53 & 0.52 \\
         0.89 & precision & 0.74 & 0.77 & 0.74 & 0.77 & 0.77 & 0.76 \\
              & ${F_1}$ & 0.59 & 0.63 & 0.60 & 0.62 & 0.62 & 0.61 \\
         \bottomrule
    \end{tabular}
     \caption{Speaker change detection results using different thresholds and time range +/-2s. The recall, precision, and ${F_1}$ of Whisper is 0.85, 0.69, and 0.76. Those of EEND is 0.60, 0.88, and 0.71, respectively.}
    \label{tab:scd}
\end{table*}

\begin{table*}[]
    \centering
    \begin{tabular}{c | c c c c c | c}
        \toprule
           language & EN-DE & EN-ES & EN-HI & EN-IT & EN-RU & Avg. \\
        \midrule
         accuracy & 0.989 & 0.989 & 0.989 & 0.989 & 0.989 & 0.99 \\
         \bottomrule
    \end{tabular}
    \caption{Token-level gender classification accuracy for different languages.}
    \label{tab:gc}
\end{table*}

\subsection{Speaker change detection}
\label{ssec:eval_scd}
A sample translation result with speaker change labels and the corresponding timestamps for text space speaker change detection is shown in Table \ref{tab:output_sentences}. Since the output is not deterministic, it is difficult to design a reference speaker change metric for translation output. Therefore, we only present a sample translation result with the speaker change label and the corresponding timestamp in Table \ref{tab:output_sentences}. We can see that the speaker change label $\langle\mathrm{SC}\rangle$ is correctly assigned to different translation results using the proposed method. The exact values of the timestamps vary slightly, but the differences are small. In fact, the difference between EN-DE and others is 1 frame. The estimated timestamps corresponding to the $\langle\mathrm{SC}\rangle$ tokens in both Table \ref{tab:output_sentences} and \ref{tab:scd} are calculated by finding the first frame generating $\langle\mathrm{SC}\rangle$ in the best hypothesis, and multiplying the frame index with frame shift, which is 0.04 second (sec) in our study.

\begin{table}[!ht]
    \setlength{\tabcolsep}{2pt}
    \centering
    \caption{A sample translation output with speaker change labels $\langle\mathrm{SC}\rangle$. 
    % The first utterance is from a male speaker and the content in English is ``It is obviously unnecessary for us to point out how luminous these criticisms are, how delicate in expression.'' The second utterance is from a female speaker and the content is ``Mister Morton then made a careful memorandum of the various particulars of Waverley's interview with Donald Bean Lean and the other circumstances which he had communicated.''. 
    % The first utterance is from a male speaker and the second is from a female speaker.
    The ground truth speaker change time is at about 8.7 sec. EN-RU and EN-HI have the same $\langle\mathrm{SC}\rangle$ locations and timestamps as those of EN-ES and EN-IT.}
    % The cosine threshold for speaker change detection is 0.94 for all the samples below except EN-RU, whose threshold is 0.77.}
    \label{tab:output_sentences}
    \begin{tabular}{@{}l p{0.74\linewidth}@{}}
        % \small
        \toprule
        language & sentence \\
        \midrule
        EN-DE &  Es ist offensichtlich unnötig darauf hinweisen, wiellos diese Kritik, wie zart im Ausdruck.\\
              & $\langle\mathrm{SC}\rangle$	8.68 sec\\
              & Herr Morton machte dann ein sorgfältiges Memorandum über die verschiedenen Einzelheiten von Waverlys Interview mit Donald Beam Lane. ...\\
              % Und die anderen Umstände, die er kommunizierte. \\
        \midrule
        EN-ES &   Obviamente es innecesario para nosotros señalar cuán brillante son estas críticas, cuán delicadas son expresión.\\
              & $\langle\mathrm{SC}\rangle$	8.72 sec\\
              & Sr. Morton luego hizo un memorándum cuidadoso de los diversos detalles de la entrevista de Waverly con Donald Beam Lane. ...\\
              % Y las otras circunstancias que había comunicado.\\
        \midrule
        % EN-HI &  "यह स्पष्ट रूप से हमारे लिए अनावश्यक है कि अभिव्यक्ति कितनी नाजुक हैं।" \\
              % & $\langle\mathrm{SC}\rangle$	8.72sec\\
              % & { श्री मॉर्टन ने वेवरली के साक्षात्कार के साथ डोनाल्ड बीम के साथ साक्षात्कार की एक सावधानीपूर्वक याद दिलाया। और अन्य परिस्थितियों में उन्होंने संवाद किया था।} \\
        EN-IT &  Ovviamente è inutile per noi sottolineare quanto siano luminosi queste critiche, quanto siano delicate un'espressione.\\
              & $\langle\mathrm{SC}\rangle$	8.72 sec\\
              & Il signor Morton poi ha fatto un attento memorandum dei vari particolari dell'intervista di Waverly con Donald Beam Lean. ...\\
              % E le altre circostanze che aveva comunicato.\\
        % \midrule
        % EN-RU & \selectlanguage{russian}
        %     Очевидно, что мы отметили, насколько ярко эти критики, насколько нежны выражения.\\
        %       & $\langle\mathrm{SC}\rangle$	8.72sec\\
        %       & \selectlanguage{russian} Затем мистер Мортон сделал тщательный меморандум о различных частях интервью Уэверли с Дональдом Бэмом. ...\\
        %       % И другие обстоятельства, о которых он сообщил.\\
        \bottomrule
    \end{tabular}
\end{table}

In addition to text space speaker change detection above, we also perform audio space speaker change detection. We use three metrics to measure the performance of speaker change detection in the time domain: recall, precision, and ${F_1}$ score. 
% Since our evaluation is based on the time domain, which has continuous values, we need a special algorithm for this task. Algorithm 1 shows how we do it. 
We define a tolerance time range for each speaker change estimation. A detected speaker change is correct if it is within the time range of the reference speaker change timestamp.

Table \ref{tab:scd} shows the recall, precision, and $F_1$ scores of the proposed method with different thresholds. The $F_1$ scores are all above 0.6 for the three thresholds of 0.99, 0.94, and 0.89. As the threshold increases, the recall improves but the precision declines. This is expected because the cosine similarity values are more likely to be below the threshold. Our test data contains noise and silence, which are challenging for the model to handle since it is not trained with such type of data. We observed that most incorrect speaker change detection points occur at noisy or silent points.
The proposed method is streaming with a chunk size of 1s, which means that speaker change detection can be estimated with very low latency. Moreover, since speaker change detection on translated text is also required for the subsequent TTS module, it is hard for traditional methods such as Whisper~\cite{radford2023robust} and EEND~\cite{Plaquet2023} to align the speaker change time with text output. Therefore, these methods are not comparable with the proposed t-vector-based method, which operates directly in token space. For comparison, we also computed the metrics for two offline models. Whisper got recall: $0.85$, precision: $0.69$ and ${F_1}$: $0.76$. EEND obtained recall: $0.60$, precision: $0.88$ and ${F_1}$: $0.71$.
Our streaming speaker change detection model achieved similar ${F_1}$ scores as the two offline models. In addition, unlike Whisper and EEND, the proposed method does not need additional steps to align the speaker change timestamp and the translated text.

\subsection{Gender classification}
\label{ssec:eval_gc}
% We achieved high accuracy by using the cosine similarity values, without needing more complex methods such as $k$-nearest neighbors with $k>1$. Moreover, many methods work in the time domain and need extra steps to match the variable ST output. Hence, we did not include those methods in our comparison.
As shown in Table \ref{tab:gc}, the token-level gender classification accuracy, which also accounts for punctuation marks, was 0.989 across different languages. This indicates that the proposed method is robust to output languages and has a high accuracy for gender classification.

\section{Conclusions}
\label{sec:conc}
This paper presents a novel challenge and a solution for streaming multi-talker ST. We combine a transducer-based multilingual ST system with a t-vector module that can identify speaker changes and gender in real time. By comparing the cosine similarity between t-vectors, our method can effectively address the speaker change detection and gender classification problems.

\bibliography{custom}

\begin{thebibliography}{45}
\providecommand{\natexlab}[1]{#1}

\bibitem[{Arivazhagan et~al.(2019)Arivazhagan, Cherry, Macherey, Chiu, Yavuz, Pang, Li, and Raffel}]{arivazhagan2019monotonic}
Naveen Arivazhagan, Colin Cherry, Wolfgang Macherey, Chung-Cheng Chiu, Semih Yavuz, Ruoming Pang, Wei Li, and Colin Raffel. 2019.
\newblock Monotonic infinite lookback attention for simultaneous machine translation.
\newblock In \emph{Proceedings of the Annual Meeting of the Association for Computational Linguistics}, pages 1313--1323.

\bibitem[{Berard et~al.(2016)Berard, Pietquin, Servan, and Besacier}]{Berard2016ST}
A.~Berard, O.~Pietquin, C.~Servan, and L.~Besacier. 2016.
\newblock Listen and translate: A proof of concept for end-to-end speech-to-text translation.
\newblock In \emph{NIPS Workshop on End-to-end Learning for Speech and Audio Processing}.

\bibitem[{B{\'e}rard et~al.(2018)B{\'e}rard, Besacier, Kocabiyikoglu, and Pietquin}]{Berard2018ST}
Alexandre B{\'e}rard, Laurent Besacier, Ali~Can Kocabiyikoglu, and Olivier Pietquin. 2018.
\newblock End-to-end automatic speech translation of audiobooks.
\newblock In \emph{IEEE International Conference on Acoustics, Speech and Signal Processing}, pages 6224--6228. IEEE.

\bibitem[{Chan et~al.(2015)Chan, Jaitly, Le, and Vinyals}]{chan2015listen}
William Chan, Navdeep Jaitly, Quoc~V Le, and Oriol Vinyals. 2015.
\newblock Listen, attend and spell.
\newblock \emph{arXiv preprint arXiv:1508.01211}.

\bibitem[{Chen et~al.(2021)Chen, Wu, Wang, Liu, and Li}]{xiechen2021tt}
X.~Chen, Y.~Wu, Z.~Wang, S.~Liu, and J.~Li. 2021.
\newblock Developing real-time streaming transformer transducer for speech recognition on large-scale dataset.
\newblock In \emph{Proc. of ICASSP}, pages 5904--5908. IEEE.

\bibitem[{Chiu and Raffel(2018)}]{chiu2018monotonic}
C.~C. Chiu and C.~Raffel. 2018.
\newblock Monotonic chunkwise attention.
\newblock In \emph{ICLR}.

\bibitem[{Fujita et~al.(2019{\natexlab{a}})Fujita, Kanda, Horiguchi, Nagamatsu, and Watanabe}]{fujita2019end}
Yusuke Fujita, Naoyuki Kanda, Shota Horiguchi, Kenji Nagamatsu, and Shinji Watanabe. 2019{\natexlab{a}}.
\newblock End-to-end neural speaker diarization with permutation-free objectives.
\newblock \emph{arXiv preprint arXiv:1909.05952}.

\bibitem[{Fujita et~al.(2019{\natexlab{b}})Fujita, Kanda, Horiguchi, Xue, Nagamatsu, and Watanabe}]{fujita2019end2}
Yusuke Fujita, Naoyuki Kanda, Shota Horiguchi, Yawen Xue, Kenji Nagamatsu, and Shinji Watanabe. 2019{\natexlab{b}}.
\newblock End-to-end neural speaker diarization with self-attention.
\newblock In \emph{2019 IEEE Automatic Speech Recognition and Understanding Workshop (ASRU)}, pages 296--303. IEEE.

\bibitem[{Graves(2012)}]{Graves-RNNSeqTransduction}
A.~Graves. 2012.
\newblock Sequence transduction with recurrent neural networks.
\newblock \emph{arXiv preprint arXiv:1211.3711}.

\bibitem[{Inaguma et~al.(2020)Inaguma, Gaur, Lu, Li, and Gong}]{inaguma2020minimum}
Hirofumi Inaguma, Yashesh Gaur, Liang Lu, Jinyu Li, and Yifan Gong. 2020.
\newblock Minimum latency training strategies for streaming sequence-to-sequence asr.
\newblock In \emph{IEEE International Conference on Acoustics, Speech and Signal Processing}, pages 6064--6068. IEEE.

\bibitem[{Kanda et~al.(2022{\natexlab{a}})Kanda, Wu, Wu, Xiao, Meng, Wang, Gaur, Chen, Li, and Yoshioka}]{kanda22arxiv}
Naoyuki Kanda, Jian Wu, Yu~Wu, Xiong Xiao, Zhong Meng, Xiaofei Wang, Yashesh Gaur, Zhuo Chen, Jinyu Li, and Takuya Yoshioka. 2022{\natexlab{a}}.
\newblock Streaming multi-talker {ASR} with token-level serialized output training.
\newblock In \emph{Proc. Interspeech}, pages 3774--3778.

\bibitem[{Kanda et~al.(2022{\natexlab{b}})Kanda, Wu, Wu, Xiao, Meng, Wang, Gaur, Chen, Li, and Yoshioka}]{kanda2022streaming}
Naoyuki Kanda, Jian Wu, Yu~Wu, Xiong Xiao, Zhong Meng, Xiaofei Wang, Yashesh Gaur, Zhuo Chen, Jinyu Li, and Takuya Yoshioka. 2022{\natexlab{b}}.
\newblock Streaming speaker-attributed asr with token-level speaker embeddings.
\newblock \emph{arXiv preprint arXiv:2203.16685}.

\bibitem[{Li et~al.(2020)Li, Zhao, Meng, Liu, Wei, Parthasarathy, Mazalov, Wang, He, Zhao, and et~al}]{li2020asr}
J.~Li, R.~Zhao, Z.~Meng, Y.~Liu, W.~Wei, S.~Parthasarathy, V.~Mazalov, Z.~Wang, L.~He, S.~Zhao, and et~al. 2020.
\newblock Developing rnnt models surpassing high-performance hybrid models with customization capability.
\newblock In \emph{Proceedings of Interspeech}, pages 3590--3594.

\bibitem[{Li(2022)}]{li2022recent}
Jinyu Li. 2022.
\newblock Recent advances in end-to-end automatic speech recognition.
\newblock \emph{APSIPA Transactions on Signal and Information Processing}, 11(1).

\bibitem[{Liu et~al.(2021)Liu, Du, Li, Li, and Chen}]{liu2021caat}
D.~Liu, M.~Du, X.~Li, Y.~Li, and E.~Chen. 2021.
\newblock Cross attention augmented transducer networks for simultaneous translation.
\newblock In \emph{Proceedings of EMNLP}, pages 39--55.

\bibitem[{Ma et~al.(2019)Ma, Pino, Cross, Puzon, and Gu}]{ma2019monotonic}
Xutai Ma, Juan~Miguel Pino, James Cross, Liezl Puzon, and Jiatao Gu. 2019.
\newblock Monotonic multihead attention.
\newblock In \emph{Proceedings of International Conference on Learning Representations}.

\bibitem[{Ma et~al.(2021)Ma, Wang, Dousti, Koehn, and Pino}]{ma2021streaming}
Xutai Ma, Yongqiang Wang, Mohammad~Javad Dousti, Philipp Koehn, and Juan Pino. 2021.
\newblock Streaming simultaneous speech translation with augmented memory transformer.
\newblock In \emph{IEEE International Conference on Acoustics, Speech and Signal Processing}, pages 7523--7527. IEEE.

\bibitem[{Matusov et~al.(2005)Matusov, Kanthak, and Ney}]{Matusov2005ST}
E.~Matusov, S.~Kanthak, and H.~Ney. 2005.
\newblock On the integration of speech recognition and statistical machine translation.
\newblock In \emph{European Conference on Speech Communicaton and Technology}.

\bibitem[{Miao et~al.(2019)Miao, Cheng, Zhang, Li, and Yan}]{miao2019online}
Haoran Miao, Gaofeng Cheng, Pengyuan Zhang, Ta~Li, and Yonghong Yan. 2019.
\newblock Online hybrid ctc/attention architecture for end-to-end speech recognition.
\newblock \emph{Proceedings of Interspeech}, pages 2623--2627.

\bibitem[{Nagrani et~al.(2020)Nagrani, Chung, Xie, and Zisserman}]{nagrani2020voxceleb}
Arsha Nagrani, Joon~Son Chung, Weidi Xie, and Andrew Zisserman. 2020.
\newblock Voxceleb: Large-scale speaker verification in the wild.
\newblock \emph{Computer Speech \& Language}, 60:101027.

\bibitem[{Nagrani et~al.(2017)Nagrani, Chung, and Zisserman}]{nagrani2017voxceleb}
Arsha Nagrani, Joon~Son Chung, and Andrew Zisserman. 2017.
\newblock Voxceleb: a large-scale speaker identification dataset.
\newblock \emph{Telephony}, 3:33--039.

\bibitem[{Ney(1999)}]{Ney1999ST}
Hermann Ney. 1999.
\newblock Speech translation: Coupling of recognition and translation.
\newblock In \emph{Proceedings of ICASSP}, pages 517--520.

\bibitem[{Papi et~al.(2024)Papi, Polak, Bojar, and Mach{\'a}{\v{c}}ek}]{papi2024real}
Sara Papi, Peter Polak, Ond{\v{r}}ej Bojar, and Dominik Mach{\'a}{\v{c}}ek. 2024.
\newblock How" real" is your real-time simultaneous speech-to-text translation system?
\newblock \emph{arXiv preprint arXiv:2412.18495}.

\bibitem[{Park et~al.(2021)Park, Kanda, Dimitriadis, Han, Watanabe, and Narayanan}]{park2021review}
Tae~Jin Park, Naoyuki Kanda, Dimitrios Dimitriadis, Kyu~J Han, Shinji Watanabe, and Shrikanth Narayanan. 2021.
\newblock A review of speaker diarization: Recent advances with deep learning.
\newblock \emph{arXiv:2101.09624}.

\bibitem[{Plaquet and Bredin(2023)}]{Plaquet2023}
Alexis Plaquet and Herv\'{e} Bredin. 2023.
\newblock Powerset multi-class cross entropy loss for neural speaker diarization.
\newblock In \emph{Proc. Interspeech 2023}.

\bibitem[{Post et~al.(2013)Post, Kumar, Lopez, Karakos, Callison-Burch, and Khudanpur}]{Post2013ST}
M.~Post, G.~Kumar, A.~Lopez, D.~Karakos, C.~Callison-Burch, and S.~Khudanpur. 2013.
\newblock Improved speech-to-text translation with the fisher and callhome spanish-english speech translation corpus.
\newblock In \emph{Proceedings of IWSLT}.

\bibitem[{Prabhavalkar et~al.(2017)Prabhavalkar, Rao, Sainath, Li, Johnson, and Jaitly}]{prabhavalkar2017asr}
R.~Prabhavalkar, K.~Rao, T.~N. Sainath, B.~Li, L.~Johnson, and N.~Jaitly. 2017.
\newblock A comparison of sequence-to-sequence models for speech recognition.
\newblock In \emph{Proceedings of Interspeech}, pages 939--943.

\bibitem[{Radford et~al.(2023)Radford, Kim, Xu, Brockman, McLeavey, and Sutskever}]{radford2023robust}
Alec Radford, Jong~Wook Kim, Tao Xu, Greg Brockman, Christine McLeavey, and Ilya Sutskever. 2023.
\newblock Robust speech recognition via large-scale weak supervision.
\newblock In \emph{International Conference on Machine Learning}, pages 28492--28518. PMLR.

\bibitem[{Sainath et~al.(2020)Sainath, He, Li, Narayanan, Pang, Bruguier, Chang, Li, Alvarez, Chen, and et~al}]{sainath2020asr}
T.~N. Sainath, Y.~He, B.~Li, A.~Narayanan, R.~Pang, A.~Bruguier, S.-Y. Chang, W.~Li, R.~Alvarez, Z.~Chen, and et~al. 2020.
\newblock A streaming on-device end-to-end model surpassing server-side conventional model quality and latency.
\newblock In \emph{Proceedings of ICASSP}, pages 6059--6003.

\bibitem[{Saon et~al.(2021)Saon, Tüske, Bolanos, and Kingsbury}]{saon2021asr}
G.~Saon, Z.~Tüske, D.~Bolanos, and B.~Kingsbury. 2021.
\newblock Advancing rnn transducer technology for speech recognition.
\newblock In \emph{Proceedings of ICASSP}, pages 5654--5658.

\bibitem[{Sperber and Paulik(2020)}]{sperber2020speech}
Matthias Sperber and Matthias Paulik. 2020.
\newblock Speech translation and the end-to-end promise: Taking stock of where we are.
\newblock In \emph{Proceedings of the 58th Annual Meeting of the Association for Computational Linguistics}, pages 7409--7421.

\bibitem[{Tang et~al.(2023)Tang, Sun, Inaguma, Chen, Dong, Ma, Tomasello, and Pino}]{tang2023hybrid}
Yun Tang, Anna~Y Sun, Hirofumi Inaguma, Xinyue Chen, Ning Dong, Xutai Ma, Paden~D Tomasello, and Juan Pino. 2023.
\newblock Hybrid transducer and attention based encoder-decoder modeling for speech-to-text tasks.
\newblock \emph{arXiv preprint arXiv:2305.03101}.

\bibitem[{Vila et~al.(2018)Vila, Escolano, Fonollosa, and Costa-Jussa}]{vila2018end}
Laura~Cross Vila, Carlos Escolano, Jos{\'e}~AR Fonollosa, and Marta~R Costa-Jussa. 2018.
\newblock End-to-end speech translation with the transformer.
\newblock In \emph{Proceedings of Interspeech}, pages 60--63.

\bibitem[{Wang et~al.(2020)Wang, Chen, Wang, Li, and Gong}]{wang2020speaker}
P.~Wang, Z.~Chen, D.~L. Wang, J.~Li, and Y.~Gong. 2020.
\newblock Speaker separation using speaker inventories and estimated speech.
\newblock \emph{IEEE/ACM TASLP}, 29:537--546.

\bibitem[{Wang et~al.(2019{\natexlab{a}})Wang, Chen, Xiao, Meng, Yoshioka, Zhou, Lu, and Li}]{wang2019speech}
P.~Wang, Z.~Chen, X.~Xiao, Z.~Meng, T.~Yoshioka, T.~Zhou, L.~Lu, and J.~Li. 2019{\natexlab{a}}.
\newblock Speech separation using speaker inventory.
\newblock In \emph{Proc. of ASRU}, pages 230--236. IEEE.

\bibitem[{Wang et~al.(2019{\natexlab{b}})Wang, Cui, Weng, and Yu}]{wang2019large}
P.~Wang, J.~Cui, C.~Weng, and D.~Yu. 2019{\natexlab{b}}.
\newblock Large margin training for attention based end-to-end speech recognition.
\newblock In \emph{Proc. of INTERSPEECH}, pages 246--250.

\bibitem[{Wang et~al.(2019{\natexlab{c}})Wang, Cui, Weng, and Yu}]{wang2019token}
P.~Wang, J.~Cui, C.~Weng, and D.~Yu. 2019{\natexlab{c}}.
\newblock Token-wise training for attention based end-to-end speech recognition.
\newblock In \emph{Proc. of ICASSP}, pages 6276--6280.

\bibitem[{Wang and Wang(2018)}]{wang2018utterance}
P.~Wang and D.~L. Wang. 2018.
\newblock Utterance-wise recurrent dropout and iterative speaker adaptation for robust monaural speech recognition.
\newblock In \emph{Proc. of ICASSP}, pages 4814--4818. IEEE.

\bibitem[{Wang et~al.(2021)Wang, Sainath, and Weiss}]{wang21t_interspeech}
Peidong Wang, Tara~N. Sainath, and Ron~J. Weiss. 2021.
\newblock \href {https://doi.org/10.21437/Interspeech.2021-683} {Multitask training with text data for end-to-end speech recognition}.
\newblock In \emph{Proc. of Interspeech}, pages 2566--2570.

\bibitem[{Wang et~al.(2022)Wang, Sun, Xue, Wu, Zhou, Gaur, Liu, and Li}]{wang2022lamassu}
Peidong Wang, Eric Sun, Jian Xue, Yu~Wu, Long Zhou, Yashesh Gaur, Shujie Liu, and Jinyu Li. 2022.
\newblock Lamassu: Streaming language-agnostic multilingual speech recognition and translation using neural transducers.
\newblock \emph{arXiv preprint arXiv:2211.02809}.

\bibitem[{Weiss et~al.(2017)Weiss, Chorowski, Jaitly, Wu, and Chen}]{weiss2017sequence}
R.~J. Weiss, J.~Chorowski, N.~Jaitly, Y.~Wu, and Z.~Chen. 2017.
\newblock Sequence-to-sequence models can directly translate foreign speech.
\newblock In \emph{Proc. of INTERSPEECH}, pages 2625--2629.

\bibitem[{Xue et~al.(2022)Xue, Wang, Li, Post, and Gaur}]{xue2022large}
Jian Xue, Peidong Wang, Jinyu Li, Matt Post, and Yashesh Gaur. 2022.
\newblock Large-scale streaming end-to-end speech translation with neural transducers.
\newblock \emph{arXiv preprint arXiv:2204.05352}.

\bibitem[{Xue et~al.(2023)Xue, Wang, Li, and Sun}]{xue2023weakly}
Jian Xue, Peidong Wang, Jinyu Li, and Eric Sun. 2023.
\newblock A weakly-supervised streaming multilingual speech model with truly zero-shot capability.
\newblock In \emph{2023 IEEE Automatic Speech Recognition and Understanding Workshop (ASRU)}, pages 1--7. IEEE.

\bibitem[{Yang et~al.(2023)Yang, Kanda, Wang, Chen, Wang, Xue, Li, and Yoshioka}]{yang2023diarist}
Mu~Yang, Naoyuki Kanda, Xiaofei Wang, Junkun Chen, Peidong Wang, Jian Xue, Jinyu Li, and Takuya Yoshioka. 2023.
\newblock Diarist: Streaming speech translation with speaker diarization.
\newblock \emph{arXiv preprint arXiv:2309.08007}.

\bibitem[{Zuluaga-Gomez et~al.(2023)Zuluaga-Gomez, Huang, Niu, Paturi, Srinivasan, Mathur, Thompson, and Federico}]{zuluaga2023end}
Juan Zuluaga-Gomez, Zhaocheng Huang, Xing Niu, Rohit Paturi, Sundararajan Srinivasan, Prashant Mathur, Brian Thompson, and Marcello Federico. 2023.
\newblock End-to-end single-channel speaker-turn aware conversational speech translation.
\newblock \emph{arXiv preprint arXiv:2311.00697}.

\end{thebibliography}

% \appendix

% \section{Example Appendix}
% \label{sec:appendix}

% This is an appendix.

\end{document}